\documentclass[useAMS,usenatbib]{mn2e}
\usepackage[dvips]{graphicx}
\usepackage{subfigure}

\usepackage{natbib}

\title[From pre- to young Planetary Nebulae: radio continuum variability]{From pre- to young Planetary Nebulae: radio continuum variability}
\author[L. Cerrigone et al.]{L. Cerrigone$^{1}$\thanks{E-mail:
lcerrigone@mpifr.de}, C. Trigilio$^{2}$,  G. Umana$^{2}$, C. Buemi$^{2}$, and P. Leto$^{2}$\\
$^{1}$Max-Planck-Institut f{\"u}r Radioastronomie, Auf dem H{\"u}gel 69, 53121 Bonn, Germany \\
$^{2}$INAF-Osservatorio Astrofisico di Catania, via S. Sofia 78, 95123 Catania, Italy}
\begin{document}

\date{}

\pagerange{\pageref{firstpage}--\pageref{lastpage}} \pubyear{2010}

\maketitle

\label{firstpage}

\begin{abstract}
Searching for variability, we have observed a sample of hot post-AGB stars and young Planetary Nebulae candidates with the Very Large Array at 4.8, 8.4, and 22.4 GHz. The sources had been previously detected in the radio continuum, which is a proof that the central stars have started ionising their circumstellar envelopes and an increase in radio flux with time can be expected as a result of the progression of the ionisation front. Such a behaviour has been found in IRAS 18062+2410, whose radio modelling has allowed us to determine that its ionised mass has increased from 10$^{-4}$ to 3.3$\times10^{-4}$~M$_\odot$ in 8 years and its envelope has become optically thin at lower frequencies.

Different temporal behaviours have been found for three other sources. IRAS~17423-1755 has shown a possibly periodic pattern and an inversion of its radio spectral index, as expected from a varying stellar wind. We estimate that the radio flux  arises from a very compact region around the central star ($\sim$10$^{15}$~cm) with an electron density of $2\times10^6$~cm$^{-3}$. IRAS~22568+6141 and 17516-2525 have decreased their radio flux densities of about 10\% per year over 4 years. 

While a linear increase of the flux density with time points out to the progression of the ionisation front in the envelope, decreases as well as quasi-periodic patterns may indicate the presence of unstable stellar winds/jets or thick dusty envelopes absorbing ionising photons.
\end{abstract}


\section{Introduction}
The evolution of planetary nebulae remains a challenging topic in
astrophysics. Planetary Nebulae (PN) have been observed over a wide
wavelength range, from X-ray to radio frequencies. Their complex
morphologies and the shaping mechanisms that produce them are still a
matter of debate.  Companion stars, jets from central stars, magnetic
fields, dust tori, and interacting winds are some of the possible shaping 
agents suggested as being responsible for various PN morphologies, 
and an overlap of their actions cannot be ruled out \citep{kwok2000}.

In general, the current theory of PN evolution is
based on the Interacting Stellar Wind (ISW) model \citep{kwok1978} and its
generalised version \citep{kahn}. The ISW model encompasses a high-velocity stellar wind driven by the hot central star that runs into the lower-velocity expanding circumstellar envelope (CSE) remaining from the earlier Asymptotic Giant Branch (AGB) phase. However, this model does not account for the shaping, because it assumes that an asymmetric distribution of matter is
already present when the wind interaction occurs. Other models take into
account the possible role of jets, and have been
successfully applied to some nebulae \citep{sahai98}. It has been
suggested that the interaction with a companion object, even a massive
planet, may provide the necessary asymmetry
\citep{soker06}. Also, large scale magnetic fields, influencing or
determining the shapes, might be sustained by a dynamo process
\citep{blackman}.

The birth of a PN is defined by the formation of an ionisation front,
which is itself a shaping agent and could heavily influence the morphology
established in earlier evolutionary phases.
In this context, important information can be provided by observations of
very young PN and pre-PN with hot central stars, where the physical processes associated with PN
formation are still occurring. The term \lq\lq young PN\rq\rq~is often used in the literature
for ne\-bu\-lae like NGC~7027, which hosts a central star with T$_\mathrm{eff} \sim2\times10^5$~K. When we use this term in this work, we actually indicate much cooler objects, which show in their optical spectra only recombination and low-excitation emission lines (see for example the spectra of some of our targets in \citet{suarez}).

To investigate the properties of these rare
objects in transition from the post-AGB to the PN, we have selected a sample
of pre-PN and young-PN candidates and searched for radio emission from ionised shells (\citet{umana}; \citet{cerrigone}).
The targets were selected from stars classified in the literature as hot post-AGB
candidates, showing strong far-IR excess and B spectral type features. 
The detection of radio continuum emission is a proof of the
presence of free electrons and therefore of ionisation. Multi-frequency observations
allow for a characterisation of the origin of the emission (i.e., different spectral indices 
point to different physical conditions).

The progression of ionisation in the envelope around the star can be observed as an increase in ionised mass
with time, which implies an increasing radio flux density. A few objects have been observed frequently enough
to allow for such an investigation. For example, CRL~618 has been observed for many years at cm and mm wavelengths and 
appears to show time intervals of increasing and decreasing flux density \citep{sanchez}. SAO~244567 is another transition object that shows a decrease in
radio flux density over a few years \citep{sao} instead of an increase, as expected on the basis of its last 40 years of evolution
\citep{bobrowsky}. Several works monitoring this class of objects in the H$_\alpha$ indicate that
the onset of ionisation may be accompanied by variability whose origin is not clear\citep{arkhipova01}, while  variability due to stellar pulsations has been  found in post-AGB stars of F--G spectral types \citep{hrivnak}.


For a better understanding of this particular phase, we have observed again all of the targets
detected in our previous works looking for radio variability. 
Six new targets following the same selection criteria of the original sample were also observed in this work, after we detected radio emission from them in a previous observing run.

\section{Observations}
We observed our targets with the Very Large Array, operated by the National Radio Astronomy Observatory (NRAO), at 4.8, 8.4, and 22.4 GHz. The observations were scheduled dynamically and only 5 sources were observed at the highest frequency, because of weather conditions and limited time available in the dynamic queue in our RA range.

The runs at the lower frequencies (4.8 and 8.4) were performed in June 2009 in the CnB and C configurations, and during reconfiguration time. 
Each target was observed for about 10 min, preceded and followed by 2 min on a nearby phase calibrator. The absolute flux density scale was defined by observing 3C48 or 3C286, depending on which of the two calibrators was observable when the run was carried out. The same flux calibrators were used in the 22.4 GHz runs, carried out in September (DnC array) and December (D array) 2009. At the highest frequency, fast switching between the target and the phase calibrator was implemented to allow for a proper phase calibration. The on-source time was also about 10 min.

The data were reduced with the \textbf{A}stronomical \textbf{I}mage
\textbf{P}rocessing \textbf{S}ystem (AIPS), according to the recommended
reduction process: the data sets were FILLMed into AIPS with average opacity
correction and nominal sensitivities (\textit{doweight}=1), edited to flag
interference or any other bad points and CALIBrated along with point
weights (\textit{docalib}=2).
Observations performed in different days were reduced separately and antenna positions were corrected with the task VLANT. Since both VLA and EVLA antennas were included in the array, 
a baseline-dependent calibration was performed for each calibrator (task BLCAL) and the solutions were then applied in the usual calibration process, to avoid excessive closure errors. Models of both 3C48 and 3C286 were used at all frequencies to properly set the absolute flux density scale.
Maps were obtained using the task IMAGR with
natural weights and CLEANed  performing a few hundreds iterations of the CLEAN algorithm.  

The flux density for each source  was estimated by fitting a
Gaussian to the unresolved source (task JMFIT), and the rms noise
was calculated in an area much larger than the synthesised beam ($>$100
beam), without evident sources in it (task IMEAN). One source (IRAS~22568+6141) was resolved in two close peaks.
In this case the data were first tapered to obtain one unresolved source and then a Gaussian was fitted, to estimate the total flux density. To better analyse the single peaks, at 4.8 and 22.4 GHz the Gaussian fitting was performed on maps obtained with uniform weighting (\textit{robust}=-5) to improve the angular resolutiona and distinguish the peaks. In spite of the uniform weighting, at 22.4 GHz, the peaks overlap; in this case the task JMFIT has been used to simultaneously fit two Gaussians. The Gaussians thus obtained are centered on the same coordinates as those at the other frequencies within about 0$''$.2, which confirms that the task correctly identifies the components.

\section{Results}
We summarise in Table~\ref{results} the flux densities obtained and their relative variation per year. The errors reported in the table contain a 3\% absolute calibration error, $\sigma=\sqrt{rms^2+(0.03 F_\nu)^2}$.
We have compared these new results with the values reported in \citet{umana} and \citet{cerrigone} looking for variability.
About half of the targets exhibit stable radio fluxes and only one is clearly variable:  IRAS 18062+2410. As \lq\lq clearly variable\rq\rq, we mean that  at both  4.8 and  8.4 GHz  we observe a relative variation  larger than three times its error. Variability in at least one band is observed in six targets, if this criterion is loosen to twice the error. Since the variations are only in one band, it is unclear whether this is due to observational errors or an intrinsic change in the emission.

For IRAS~22568+6141, which is resolved into two blobs of emission, we list in Table~\ref{results} the flux densities of its Northern and Southern lobe as well as the emission from the whole nebula. As explained in the previous section, the flux density of the whole nebula was obtained by fitting a Gaussian on a tapered map, where the source is not resolved. The sum of the flux densities of the two  blobs does not account for the emission of the entire nebula. An inspection of the residual maps (the maps after subtraction of the fitting Gaussians) indicates that about 4.3~mJy at 4.8 GHz and 0.5~mJy at 8.4 GHz remain unmatched. This means that - besides the two peaks - weak extended emission is present and is recovered by tapering.

\begin{table*}
\centering
\caption{Radio flux densities obtained at the VLA and their relative variations per year. Targets that do not show variations are listed first, then targets that show variations at least in one band. For IRAS 22568+6141, the flux densities of its Northern and Southern lobes are also listed.}
\label{results}
\begin{tabular}{@{}lccccc}
\hline
IRAS ID	&	F$_\mathrm{4.8\, GHz}$	&	$\Delta \mathrm{F_{4.8\,GHz}}$ &	F$_\mathrm{8.4\, GHz}$	& $\Delta \mathrm{F_{8.4\,GHz}}$ &	F$_\mathrm{22.4\, GHz}$	 \\
			&	mJy		&  \% yr$^{-1}$	& mJy			& \% yr$^{-1}$	&	mJy		\\
\hline
01005+7910		&	0.29$\pm$0.05	& --		&	0.17$\pm$0.04	& -4.3$\pm$4.9 	&	-- 	\\
17381-1616		&	1.64$\pm$0.08	& 0.3$\pm$1.2	&	1.62$\pm$0.07	&  2.3$\pm$1.4	&	--		\\
19336-0400		&	11.1$\pm$0.3	& 1.1$\pm$1.0	&	9.5$\pm$0.3	& -0.3$\pm$1.0		&	--        \\
19590-1249		&	3.3$\pm$0.1	& 1.1$\pm$1.3	&	3.1$\pm$0.1	& 1.6$\pm$0.9	&	--       \\
20462+3416		&	0.62$\pm$0.05	& 0.7$\pm$2.0	&	0.43$\pm$0.04	& 0.4$\pm$2.6	&	--          \\
21546+4721		&	1.56$\pm$0.06	& --		&	1.56$\pm$0.06	&  0.6$\pm$1.4	&	1.32$\pm$0.07      \\
22023+5249		&	2.60$\pm$0.09	& --		&	2.28$\pm$0.08	& 2.3$\pm$1.2	&	2.1$\pm$0.1      \\
22495+5134		&	9.6$\pm$0.3	& --		&	8.6$\pm$0.3	& 0.8$\pm$1.2	&	8.1$\pm$0.3     \\
			&			&		&			&		&			\\
06556+1623		&	0.45$\pm$0.05	& -4.1$\pm$1.5	&	0.48$\pm$0.04	& -2.1$\pm$1.5	&	0.11$^a$	\\
17423-1755		&	0.63$\pm$0.06	& --		&	0.50$\pm$0.05	&  15.4$\pm$4.8	&	--		\\
17460-3114		&	1.4$\pm$0.1	& -1.7$\pm$1.4	&	1.08$\pm$0.07	& -2.7$\pm$1.3	&	--        \\
17516-2525		&	0.4$\pm$0.1$^b$	& --		&	0.32$\pm$0.06	& -11.6$\pm$2.8	&	--        \\
18062+2410		&	1.72$\pm$0.08	& 4.9$\pm$1.5	&	2.67$\pm$0.09	& 13.8$\pm$2.2	&	--         \\
18442-1144		&	21.7$\pm$0.7	& 3.7$\pm$1.2	&	20.7$\pm$0.6	& 1.3$\pm$1.0	&	--      \\
22568+6141		&	23.8$\pm$0.7	& --		&	21.7$\pm$0.7	& -8.1$\pm$0.7	&	16.0$\pm$0.5   \\
22568+6141 N		&	10.0$\pm$0.3	& --		&	11.2$\pm$0.3	& --		&	8.5$\pm$0.2	  \\
22568+6141 S		&	9.8$\pm$0.3	& --		&	9.3$\pm$0.3	& --		&	7.3$\pm$0.3 \\
\hline
\end{tabular}
\flushleft
$^a$The source was not detected and the rms of the map is listed. \\
$^b$Marginal detection.
\end{table*}


We performed for the first time multi-frequency observations of IRAS 01005+7910, 17516-2525, 21546+4721, 22023+5249, 22495+5134, and 22568+6141. 
\begin{figure*}
\centering
\subfigure[]{\includegraphics[width=7.5cm]{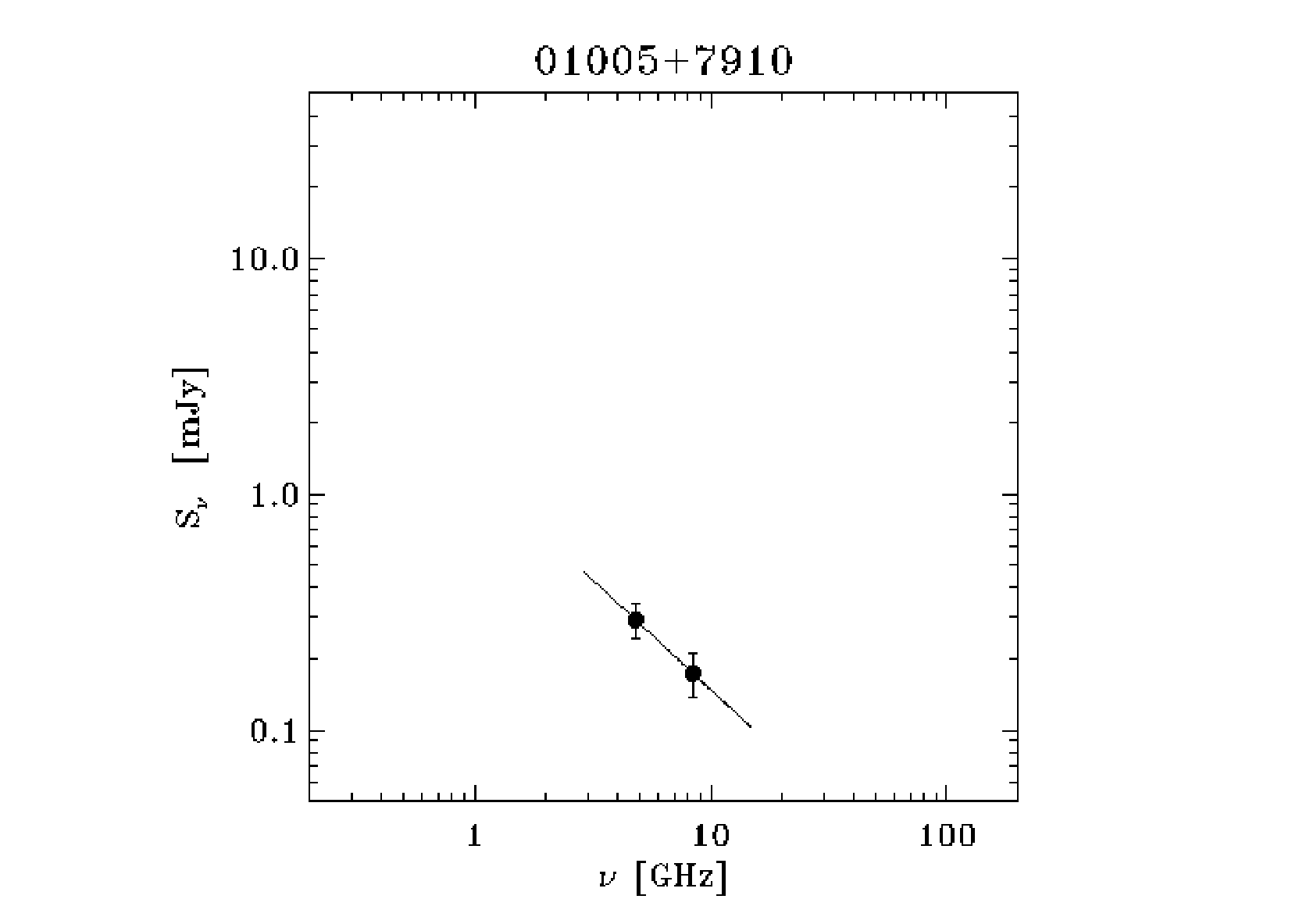}}
\subfigure[]{\includegraphics[width=7.5cm]{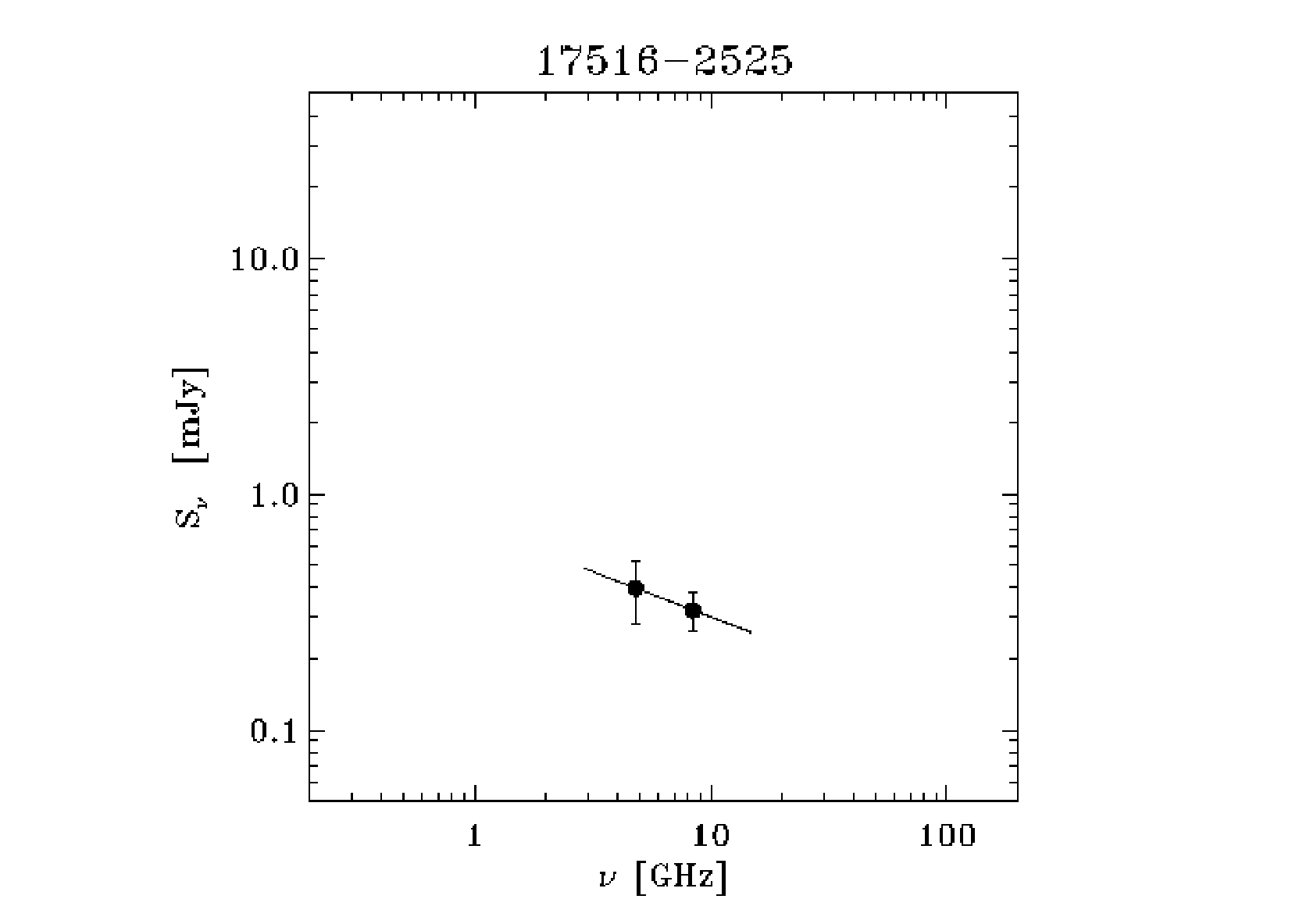}}
\subfigure[]{\includegraphics[width=7.5cm]{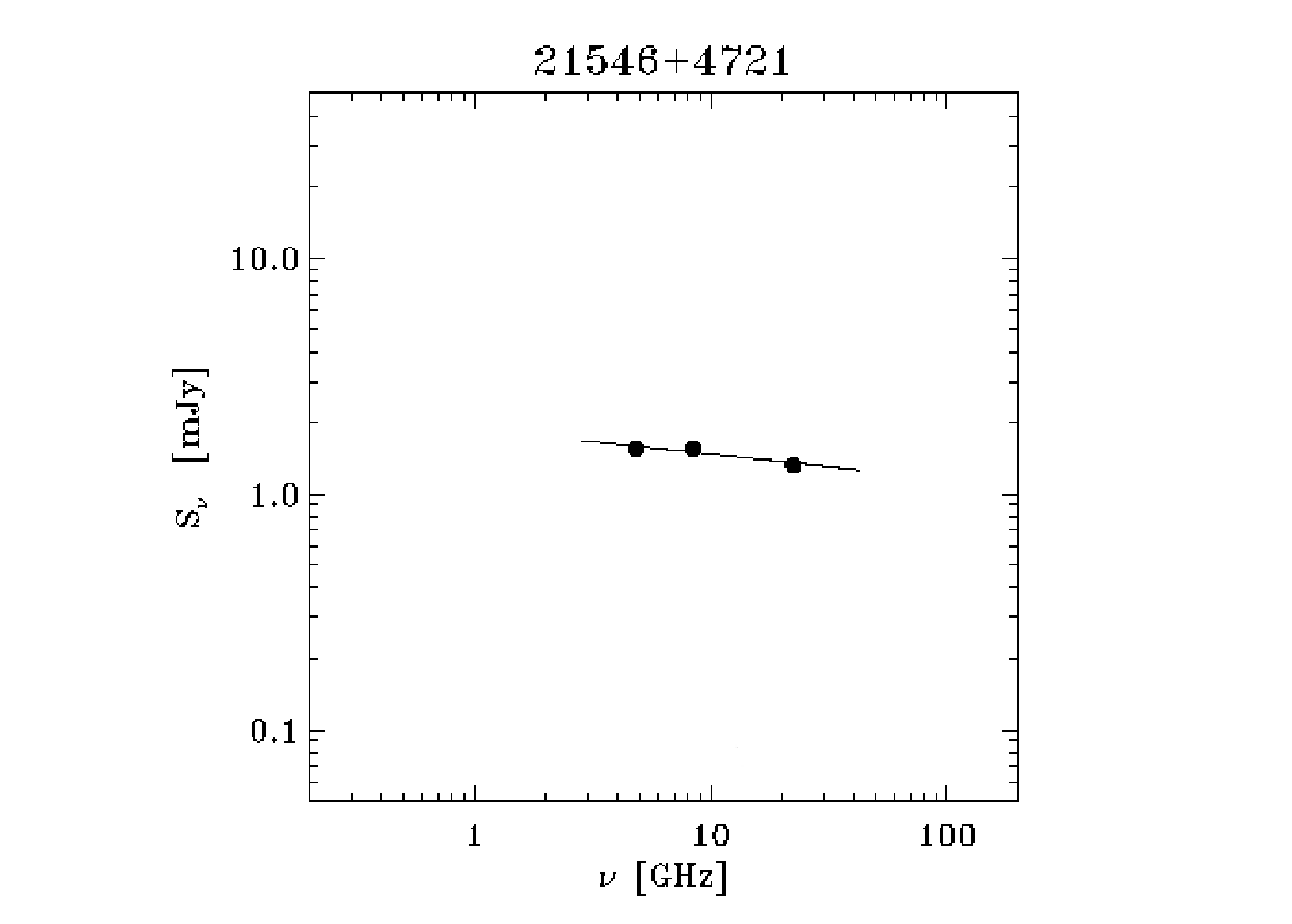}}
\subfigure[]{\includegraphics[width=7.5cm]{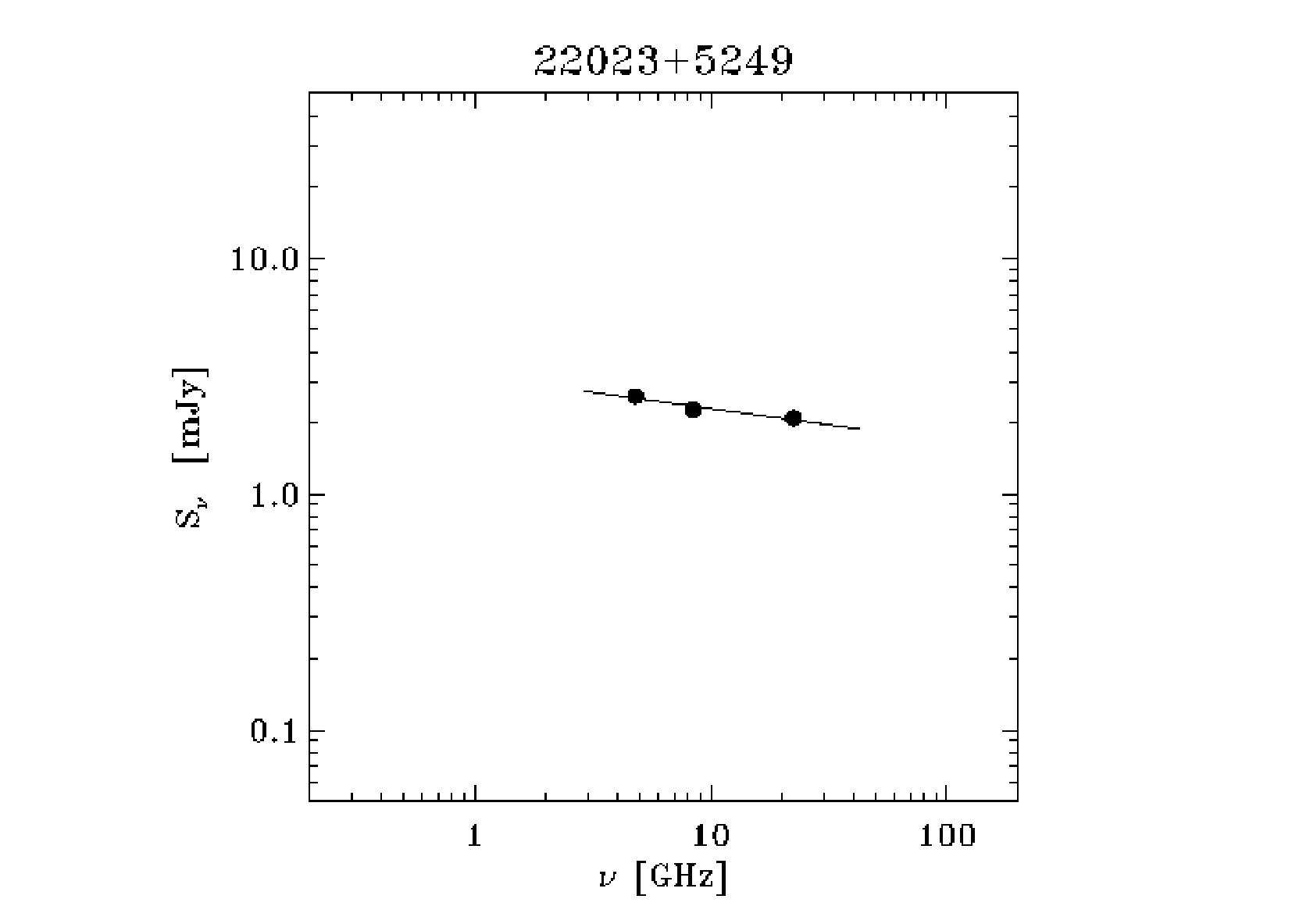}}
\subfigure[]{\includegraphics[width=7.5cm]{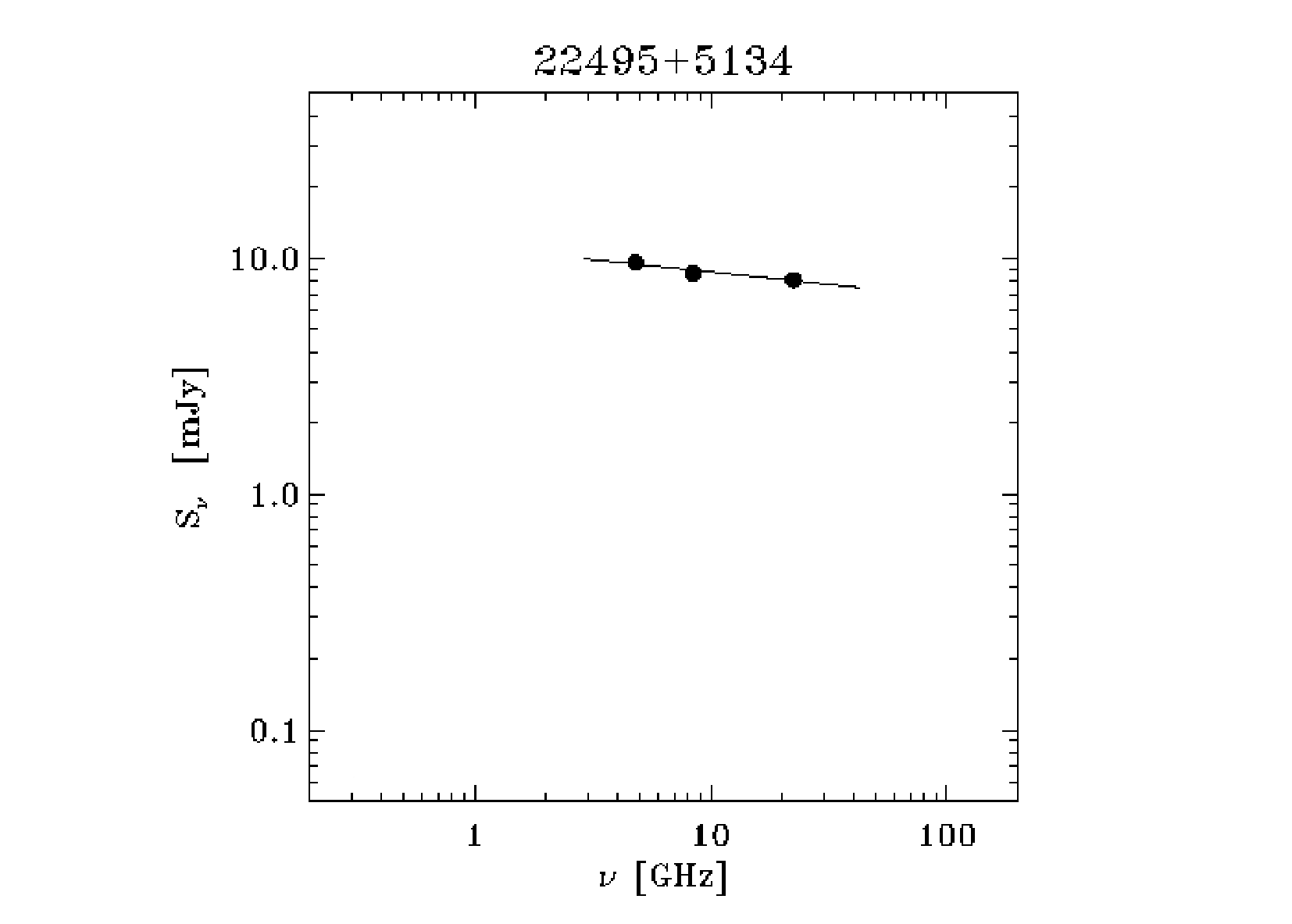}}
\subfigure[]{\includegraphics[width=7.5cm]{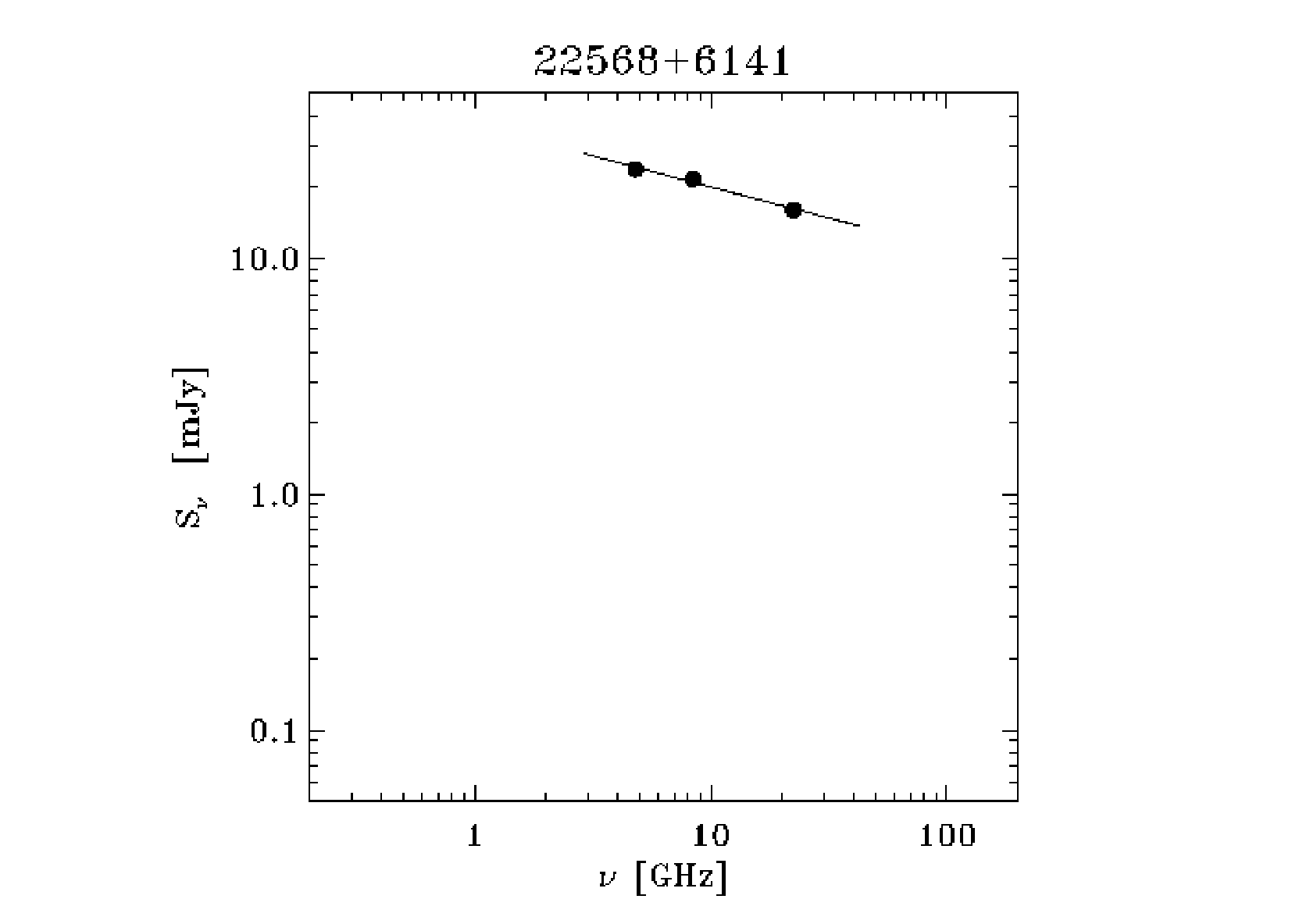}}
\caption{Continuum spectra for targets that had not been observed before at radio wavelengths or that lacked multi-frequency observations.} 
\label{new_spectra}
\end{figure*}

We fitted the data points to derive the spectral indices, which turned out to match very well with what expected for optically thin shells ($\alpha\sim$~-0.1), as can be seen in Table~\ref{fits}.
Since the emission is optically thin, we can also derive the emission measure  for each of these sources  as 
\[
<EM>=\frac{\int_\Omega EM d\Omega}{\Omega}=\frac{5.3\times10^5\,F_{8.4\,GHz}}{\theta^2}
\]
where $F_{8.4 \, GHz}$ is the flux density at 8.4~GHz in mJy, and $\theta$ the angular radius in arcsec \citep{terzian}. The only target that is resolved at our angular resolution is IRAS~22568+6141 (\S~\ref{para22568}), whose northern and southern blobs of emission are listed separately in Table~\ref{fits}. Gaussian fitting of the two peaks allows us to determine their positions as RA=22:58:51.44 DEC=61:57:45.2 (North peak) and RA=22:58:51.68 DEC=61:57:42.8 (South peak). The peaks are about 3$''$ apart, with PA$\sim$146$^\circ$. 

To calculate their emission measures, we need to estimate the angular sizes of our sources. In Table~\ref{fits}, we list the major and minor axes of the convolution beam. For those sources that have been resolved with high-angular resolution observations  in \cite{cerrigone} (IRAS~18442-1144, 19336-0400, 19590-1249, and 22023+5249), the geometric mean of the axes is an approximate estimation of their diameters, therefore we take into account this value in our calculations.


\begin{table}
\centering
\caption{Spectral indices from data fitting  (with statistical errors), emission measures (lower limit) and beam sizes of the 8.4 GHz data. The list is divided in sources previously detected (upper list) and new detections (lower list).}
\label{fits}
\begin{tabular}{@{}lccc}
\hline
Target & Spec. Index & Emission measure & Beam \\
       &             & 10$^5$~cm$^{-6}\,$pc &  arcsec$^2$  \\
\hline
06556+1623 &	 0.10$\pm$0.47 & 2.6$\pm$0.2	 & 3.1$\times$1.3 \\
17381-1616 &	-0.02$\pm$0.22 & 2.6$\pm$0.1	 & 4.7$\times$2.8 \\
17423-1755 &	-0.40$\pm$0.47 & 0.78$\pm$0.07	 & 4.9$\times$2.8 \\	
17460-3114 &	-0.51$\pm$0.35 & 1.30$\pm$0.08	 & 6.9$\times$2.6 \\	
18062+2410 &	 0.78$\pm$0.20 & --$^a$	         & 3.0$\times$2.7 \\	
18442-1144 &	-0.09$\pm$0.15 & 38.2$\pm$0.1	 & 4.2$\times$2.8 \\	
19336-0400 &	-0.27$\pm$0.15 & 20.4$\pm$0.1	 & 3.6$\times$2.7 \\	
19590-1249 &	-0.13$\pm$0.17 & 9.4$\pm$0.1	 & 3.0$\times$2.3 \\	
20462+3416 &	-0.67$\pm$0.46 & 2.0$\pm$0.2	 & 2.6$\times$1.7 \\	
           &                &	 &	\\
01005+7910 & -0.93$\pm$0.98 & 0.8$\pm$0.2   &      2.8$\times$1.7  \\
17516-2525 & -0.38$\pm$1.29 & 0.5$\pm$0.1   &      5.7$\times$2.2 \\
21546+4721 & -0.11$\pm$0.09 & 7.8$\pm$0.2   &       2.7$\times$1.6 \\
22023+5249 & -0.14$\pm$0.08 & 8.6$\pm$0.2  &       2.6$\times$2.1 \\ 
22495+5134 & -0.11$\pm$0.06 & 11.7$\pm$0.2  &       5.3$\times$2.9 \\
22568+6141 & -0.26$\pm$0.05 & -- & --  \\
22568+6141 N & -0.14$\pm$0.02 & 148$\pm$4 & 2.0$\times$0.8  \\
22568+6141 S & -0.20$\pm$0.04 &  123$\pm$4 & 2.0$\times$0.8  \\
\hline
\end{tabular}
\flushleft
$^a$~Not optically thin: see modelling in \S~\ref{18062}
\end{table}

\section{IRAS 17423-1755}
IRAS 17423-1755 (Hen3-1475) is a well known point-symmetric pre-PN at a distance of about 5.8~kpc. It shows OH maser emission, lines from ionised elements, and recombination lines due to shocks propagating in its CSE \citep{riera}.  \citet{sanchez01} find that two wind components are present in the vicinity of the star ($\sim0''.7$): a fast and an ultra-fast wind (150--1200 and 2300~km~s$^{-1}$ respectively), both with kinematical ages of tens of years. Decreasing velocities from 1000 to 150 km~s$^{-1}$ are  found in the knots in its bipolar outflow, traced by  optical and near-IR imaging over 17$''$ \citep{riera}. Millimeter and cm observations trace much smaller structures, likely associated to a circumstellar disk (millimeter observations) and a compact ionised region (centimeter observations) \citep{huggins}. 

Several models have been applied to this source to account for its kinematical and morphological properties. A possible explanation is that its pecularities are the result of an ejection variability of the central source \citep{riera}. 
\citet{velazquez} have modelled this nebula with a  precessing jet of periodically variable ejection velocity superimposed to a linear increase, which propagates into an Interstellar Medium (ISM) previously perturbed by an AGB wind. The velocity variability in their model has a period of 120~yr and a half-amplitude of 150~km~s$^{-1}$. The jet  has therefore an ejection velocity of the form $v_{jet} = v_0 + v_1 sin (2 \pi(t - t_0 ) / \tau) + a t$, with $v_0= 400$~km~s$^{-1}$, $v_1 = $150~km~s$^{-1}$, $t_0=-640 yr$, $\tau=$~120~yr, and $a = $1~km~s$^{-1}$~yr$^{-1}$.

\citet{lee} present a general model to shape pre-PN with Collimated Fast Winds (CFW) interacting with a spherical AGB wind and they compare their results with observations of CRL 618. 
They introduce periodic variations of the density and velocity of the fast wind in such a way that the mass-loss rate ($\dot{\mathrm{M}}=4\pi$r$^2\rho$v) is kept constant: the density varies as $\rho=\frac{\rho_f}{1+A \sin\frac{2\pi t}{\tau}}$ and the velocity as $v=v_f (1+A \sin\frac{2\pi t}{\tau})$.
In particular, to compare their model to CRL 618 they consider a period of 22~yr and a velocity half-amplitude of 150~km~s$^{-1}$.

\begin{figure}
\centering
\includegraphics[width=8cm]{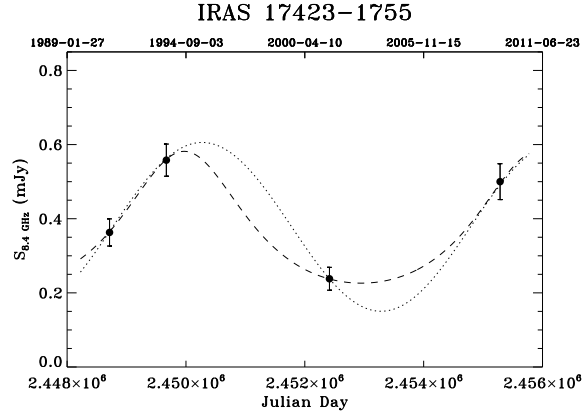}
\caption{VLA observations at 8.4 GHz of IRAS 17423-1755 over 16 years. The point in 1991 is from the VLA project AK247 (PI: Knapp), while that from 1993 is from AZ63 (PI: Zijlstra).}
\label{17423_lightcurve}
\end{figure}

Radio continuum data since 1991 are available in the VLA archive. 
We retrieved the data, reduced them following the same recipe as for the rest of our data, and found a flux density of 0.36$\pm$0.04~mJy in 1991 and 0.56$\pm$0.04~mJy in 1993, both at 8.4 GHz. 
As shown in Figure~\ref{17423_lightcurve}, the combination of our 2001 and 2009 observations with the archive data clearly shows  variability of the radio flux density. Unfortunately, our temporal sampling is too coarse to conclusively  assess whether the variability is regular or not. Nevertheless, we show that the data are well fitted by periodic functional forms similar to those introduced by \citet{velazquez} and \citet{lee}: 
\[
\begin{array}{ll}
S_\nu = {A_0\,}{\sin{\frac{2\pi}{A_1}t}}+A_2  & \qquad \qquad\qquad  \mbox{(dotted line)}\\ 
S_\nu = \frac{A_0}{1 + A_1 \sin{\frac{2\pi}{A_2}t}}  & \qquad \qquad\qquad \mbox{(dashed line)} \\
\end{array}  
\]

We are fitting four points with functions having three degrees of freedom, therefore the procedure should not be over-interpreted. Both fits give periods around 19~yr (18.6 and 18.7~yr, for the dashed and dotted curves, respectively), comparable to the periodicity in the model of CRL~618 by \citet{lee}, but different from the period introduced in the model of IRAS 17423-1755 by \citet{velazquez}.

Our 2009  observations  indicate that the emission is optically thin, with a spectral index of -0.4 (Table~\ref{fits}). Nevertheless, the observations reported in \citet{cerrigone} at 8.4 and 22.4 GHz indicate optically thick emission between these frequencies  with a spectral index around 1. In the cited paper, it was assumed that no variation had occurred in the source between 2001 (epoch of the 8.4 GHz data) and 2003 (epoch of the 22.4 GHz data). Figure~\ref{17423_lightcurve} shows that this assumption is not correct  and that in 2003 the flux density at 8.4 GHz was smaller than in 2001, if the variability is regular. This does not affect the conclusion that the spectral index was positive in 2003 and likely around 1, considering errors.

 If the radio emission is due to the stellar wind, the radius R of the emitting region depends on the frequency as R($\nu$)$\sim\nu^{-0.7}$ \citep{panafelli}, in the optically thick regime, as is the case in the 2003 observations. 
Since any variation in the wind will propagate outwards and the emission at higher frequency arises from inner regions, a light curve at a lower frequency will be delayed compared to one at a higher frequency. Therefore, the value of the spectral index will also vary with time, being positive when the emission is decreasing and negative when it is increasing. The measurement of a positive spectral index in 2003 and a negative one in 2009 strengthens then the interpretation of the observed variability as due to an unstable stellar wind.
As shown in \citet{panafelli}, if the radio flux is due to a stellar wind,  we can calculate its mass loss rate as 
\[
\dot{M}=0.32\times10^{-5} \frac{v_\infty}{10^3 \;\mathrm{km~s}^{-1}} \left(\frac{F_\nu}{\mathrm{mJy}}\right)^{3/4} \left(\frac{D}{\mathrm{kpc}}\right)^{3/2} \mathrm{M}_\odot \mathrm{yr}^{-1}
\]
and the size of the emitting region as 
\begin{eqnarray*}
R(\nu) & = & 6.23\times10^{14}\left(\frac{\nu}{10 \;\mathrm{GHz}}\right)^{-0.7}\left(\frac{\dot{M}}{10^{-5} \;\mathrm{M}_\odot \,\mathrm{yr}^{-1}}\right)^{2/3} \\
       &   & \cdot \left(\frac{v_\infty}{10^3 \;\mathrm{km~s}^{-1}}\right)^{-2/3} \;\mathrm{cm}\\
\end{eqnarray*}
where cosmic abundances, full ionisation, and a 10$^4$~K temperature of the electron gas have been assumed.
If we link the radio flux to the fast wind component, we can take 100~km~s$^{-1}$ as its terminal velocity and calculate from the radio flux density in 2001 (when the spectral index was likely positive) $\dot{\mathrm{M}}\sim 1.6\times10^{-6}$~M$_\odot$~yr$^{-1}$ and R$_{8.4 \,\mathrm{GHz}}\sim 10^{15}$~cm ($\sim0''$.01), assuming a distance of 5.8~kpc. 

The mass loss rate found is quite high for a post-AGB star, where values smaller than 10$^{-7}$ M$_\odot$~yr$^{-1}$ are expected. Nevertheless, large rates (10$^{-5}$--10$^{-4}$ M$_\odot$~yr$^{-1}$) have been estimated in the prototypical pre-PN CRL~618, which has undergone  at least two distinct episodes of mass loss in the form of a slow wind, in the last 2500 yr \citep{sanchez}. Also, enhanced mass loss in IRAS~17423-1755 has been found by \citet{huggins}, who have estimated that its molecular envelope ($\sim$0.6 M$_\odot$) was ejected in less than 1500~yr at a mass loss rate  $>10^{-4}$~M$_\odot$~yr$^{-1}$. The estimation of the linear size of the emitting region allows us to improve the calculation of the emission measure and convert it into an electron density of 2$\times10^6$~cm$^{-3}$.

If we assume that, like in CRL~618, the radio flux arises from a slow stellar wind  (15~km~s$^{-1}$), its mass loss rate becomes $2.4\times10^{-7}$~M$_\odot$~yr$^{-1}$, with R$_{8.4}$ approximately unchanged. 
 
The lack of multi-epoch observations covering the whole cm range does not allow for conclusiveness.

\section{IRAS 17516-2525}
This source was first studied by \citet{vanderveen}, who had selected it as a post-AGB candidate because of its IRAS colours. Hydroxil maser lines  with $\Delta v\sim40$~km~s$^{-1}$ were detected with the Parkes and Nan\c{c}ay radiotelescopes by \citet{telintel} and \citet{szymczak} respectively; the latter detected both 1612 and 1667 MHz lines. Nevertheless, \citet{vanderveen} did not detect the OH 1612 MHz line with the VLA in B array at the coordinates of the source, but $\sim$7$'$ north of it. It is likely then that the detections performed with the single-dish telescopes are to be associated to this off-set maser, although further interferometric observations would be desirable to rule out an intrinsic OH variability. 

\citet{sanchez08} performed  optical spectroscopy of IRAS~17516-2525 and found that its spectrum is dominated by nebular emission lines. They also suggest the presence of a hot central star, based on the similarity between its spectrum and that of IRAS~19520+2759, which they classify as O9. In their analysis of the optical spectrum, \citet{sanchez08} find that the H$\alpha$ has a P-Cyg profile, indicative of on-going mass loss, although Pa lines do not display the same profile. Furthermore, they detect the $[$CaII$]$ doublet at 7291.47 and 7323.89~\AA, which leads them to conclude that moderate-velocity (40$<$v$<$100~km~s$^{-1}$) shocks must propagate in the envelope around this star and distruct dust grains, enhancing the amount of Ca in the gas phase, which would otherwise be depleted onto grains.

We detected IRAS~17516-2525 for the first time in 2005 and find a strong decrease of its flux density over 4 years (11\% per year), the source having almost halved its emission at 8.4 GHz. The spectral index between 4.8 and 8.4 GHz indicates an optically thin spectrum, although the detection at the lower frequency is only marginal. In the context of the observations available in the literature, the radio continuum variability may be due to instabilities of the stellar wind (i.e., shocks) propagating in the circumstellar envelope. More and more sensitive observations are needed to inspect this target and determine its radio continuum spectrum, to possibly distinguish between an ionised shell and a stellar wind, and to look for a variability pattern. Within the uncertainty due to the larger error at the lower frequency, our estimate of a flat radio spectral index strengthens the hypothesis that this is a young PN.

\section{IRAS 18062+2410}
\label{18062}
This source has been monitored at optical wavelengths over a wide period of time and  shows photometric variability  on both short (hours) and long (years)  time scales. In particular, the equivalent widths of H$_\alpha$ and H$_\beta$ increase and its optical continuum decreases \citep{arkhipova}. While the short-time variability has been explained in terms of stellar pulsations and an unstable stellar wind, the long-timescale variations have been attributed to an increase of the degree of ionisation of the envelope, due to a stellar temperature increasing at a  rate of about 200~K per year \citep{arkhipova}.

Previous VLA observations of this target performed in 2003 and 2005 are reported in \citet{cerrigone}. 
We have modelled the radio continuum emssion assuming that the central
star is surrounded by a shell of ionised gas with density radially
decreasing as $r^{-2}$.  We have calculated a 3D distribution of mass in a
spherical shell with outer radius R$_{out}$, inner radius
R$_{in}=\eta$R$_{out}$ (with $0<\eta<1$), and density at the inner radius
$\rho_{in}$. The calculation of the average density in the shell links this parameter to the density at the inner radius: $<\rho>\approx\eta^{2}(1-\eta)\rho_{in}$. 
The electron temperature was set
to 10$^4$~K. Details of the radio model can be found in
\citet{sao}. We fit the 2003 data with  R$_{in}$=0$''$.0354, R$_{out}$=0$''$.06, $\rho_{in}$=5.2~10$^5$~cm$^{-3}$, and d=6.4 kpc, which gives us an estimated ionised mass about 1.48~10$^{-4}$~M$_\odot$.

\begin{figure}
\centering
\includegraphics[width=8cm,clip]{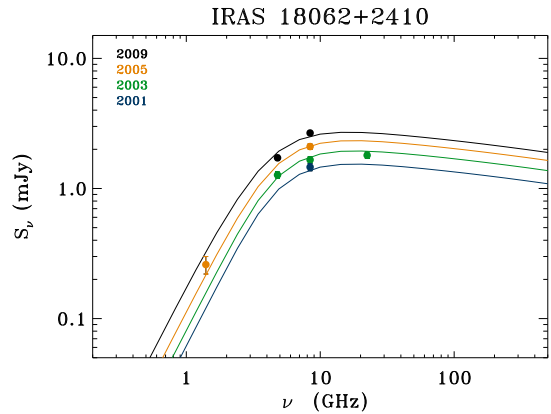}
\caption{Radio continuum spectra of IRAS 18062+2410 obtained at different epochs. The correspondent model curves are shown in the same colour as the data points.}
\label{18062_spectrum}
\end{figure}

The nebula has increased its flux density in 2005 and 2009, as shown in Figure~\ref{18062_spectrum}. To estimate the expansion velocity of the ionisation front, we assume that the CSE expands at a typical velocity of 15 km~s$^{-1}$ (therefore its inner radius will be 0$''$.0364 in 2005). A fit to the 2005 data  enables us to calculate that the velocity of the ionisation front is of the order of 100--200 km~s$^{-1}$, the  typical range for fast winds in post-AGB stars. 

Having the above mentioned values as first guesses, we now try to fit all of the 2001--2009 data keeping the CSE expansion velocity set at 15 km~s$^{-1}$. We first set R$_{in0}$, R$_{out0}$ (the values in 2001), and $\rho_{in0}$ (the density at R$_{in0}$, which does not vary with the epoch) and $v_{out}$ (the velocity of the ionisation front within the shell). Then we calculate R$_{in}$=R$_{in0}+v_{in}\Delta t$ and R$_{out}$=R$_{out0}+(v_{in}+v_{out})\Delta t$  for the other epochs, $\Delta t$ being the time between the first observation and the specific epoch; finally, we calculate the density at R$_{in}$ as $\rho_{in}$=$\rho_{in0} \left(\frac{\mathrm{R}_{in0}}{\mathrm{R}_{in}}\right)^2$. A satisfactory fit can be obtained if $v_{out}$=120 km~s$^{-1}$ (kept constant at all epochs) and in 2001 we have R$_{in}$=0$''$.0380, R$_{out}$=0$''$.0522, and $\rho_{in}$=5.38~10$^5$~cm$^{-3}$, as shown in Figure~\ref{18062_spectrum}. 

The models allow us to estimate that the critical frequency has decreased from around 8.2 GHz in 2001 to 7.6 GHz in 2009, pointing to an optical thinning of the envelope, with the opacity at 8.4 GHz ranging between 0.96 in 2001 and 0.84 in 2009. The ionised masses are  $1.0\times10^{-4}$, $1.5\times10^{-4}$, $2.2\times10^{-4}$, and $3.3\times10^{-4}$ M$_\odot$, respectively in 2001, 2003, 2005, and 2009.

 

Although a linear increase of the flux density produces a good fit to the data, it cannot rule out the possibility of a cyclic variability with a period much longer than 6 years. 

\begin{figure}
\centering
\includegraphics[width=8cm]{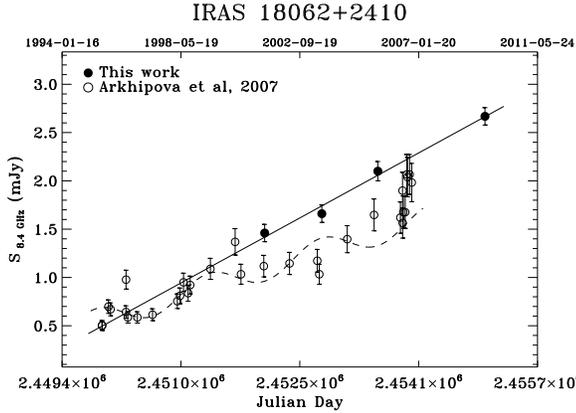}
\caption{Radio flux density at 8.4~GHz and H$_\beta$ flux \citep{arkhipova} at several epochs for IRAS 18062+2410. The dashed curve is a fit to the H$_\beta$ points before 2002.}
\label{18062_lightcurve}
\end{figure}

Figure~\ref{18062_lightcurve}, shows the variability that we find in this object, which confirms what reported in \citet{cerrigone}. We have overplotted the 8.4 GHz flux density calculated from the H$_\beta$ fluxes in \citet{arkhipova}, which we have estimated from the equivalent widths and absolute values in 1996 and 2006 given in that work. \citet{arkhipova} estimate that the H$_\beta$ flux  was 2.7$\pm$0.3~10$^{-13}$~erg~cm$^{-2}$~s$^{-1}$ in 1995-1996 and 6.2$\pm$0.4~10$^{-13}$~erg~cm$^{-2}$~s$^{-1}$ in 2006 and list its equivalent width over 10 years. To convert the equivalent width into flux, we have calculated the average width over all of the 1995-1996 observations (one single high point was neglected) and over the 2006 data, then scaled the average values to the respective fluxes given in the paper, and finally averaged these two scaling factors together to obtain one mean equivalent-width/flux conversion factor.   After converting into flux units each equivalent width, these were converted into radio flux densities at 8.4 GHz following Eq. IV-26 in \citet{pottasch}
\[
\begin{array}{l}
S_{\nu}=2.51\times10^7 \,T{_e}^{0.53}\, \nu^{-0.1}\, Y\, F_{H_\beta}
\end{array}
\]
where S$_{\nu}$ is the flux density in Jy at the frequency $\nu$, T${_e}$ is the electron temperature (set to 10$^4$ K), $\nu$ is in GHz, Y accounts for contributions from He ions and is set to 1.1, and F$_{H_\beta}$ is the flux in the H$_\beta$ in units of erg~cm$^{-2}$~s$^{-1}$.

 The mismatch between the two data sets in Figure~\ref{18062_lightcurve} is likely  due to an optical-depth effect. As we have shown in our modelling, the nebula is undergoing optical thinning of its envelope. When the radio flux density is optically thick, its observed value will be smaller than that calculated from the H$_\beta$ flux; on the contrary, when it is optically thin, it will be larger than the value calculated from the H$_\beta$, because of absorption at optical wavelengths. Figure~\ref{18062_lightcurve} seems to be showing the passage from one regime to the other. 

The H$_\beta$ data seem to show a periodicity before 2002. After excluding three clear outsider points, we have calculated a fit to the data before 2002 with a functional form of type $F_{H_\beta} = A_0 sin (2 \pi t / A_1) + A_2 t + A_3 $, which includes a periodic term and a linearly increasing one. This fit gives a period of $\sim$4.1~yr and a starting time $\sim$6.3 yr before the first observation ($A_0$=0.47, $A_1=1500$, $A_2=8.7172$~10$^{-4}$, $A_3=-2133.622$).

The number of Lyman photons necessary to sustain the ionised region can be calculated as 
\[
Q=V n_e^2 \alpha_B
\]
where $V$ is the volume of the ionised region, $n_e$ the average electron density, and $\alpha_B$ the recombination coefficient summed over all the excited states ($2.59\times10^{-13}$~cm$^3$~s$^{-1}$, at T$_e$=10$^4$~K) \citep{kwokbook}. Taking advantage of the parameters derived from our modelling, we can estimate that in 2001 $Q=2.2\times10^{46}$~photons~s$^{-1}$, while in 2009 $Q=1.0\times10^{47}$~photons~s$^{-1}$, which by interpolating the values in \citet{panagia} give temperatures of the central star of $\sim$23400 and $\sim$25900 K, in 2001 and 2009 respectively.

\section{IRAS 22568+6141}
\label{para22568}
IRAS~22568+6141 was discovered by \citet{garcialario}, who classified it as a possible young PN and estimated a distance of 6~kpc, electron density about $2\times10^4$~cm$^{-3}$ and T$_{\mathrm{eff}}\sim$24000~K. For this source, a comparison with previous data is possible only at one frequency. We first detected this nebula with the VLA (C array) in July 2005 at 8.4 GHz and found a flux density of 32.70$\pm$0.98 mJy (Umana et al, in preparation). The source was also included in a survey carried out with the Torun 32-m Radio Telescope between December 2005 and May 2007 by \citet{pazderska}. They measured 31.4$\pm$4.1~mJy at 30 GHz, which is in agreement with our measurement at 8.4 GHz, because we find that the emission is optically thin. 

Our observations indicate a decrease in the emission around 34\% between 2005 and 2009. 
Since the source is known to be extended and we observed it with an interforometer in different array configurations, missing flux could be the reason of the apparent variability. Interferometers in fact filter out emission from structures beyond a size - the Largest Angular Scale (LAS) - that depends on the shortest baseline and on the \textit{uv}-coverage. Our low-frequency observations were carried out during reconfiguration time and the minimum baselines turned out to be about 1.3 and 2.1~k$\lambda$ at 4.8 and 8.4 GHz respectively. These correspond respectively to a LAS around 40$''$ and 23$''$ in snapshot mode. The high-frequency run was performed in D array, with LAS$\sim$33$''$. It is therefore unlikely that we are missing any flux from extended structures, since this source is known to have a size around 8$''$ from observations in other wavelength ranges \citep{sahai07}. We can conclude that the fading of the nebula is real. 

\begin{figure}
\centering
\includegraphics[width=7cm]{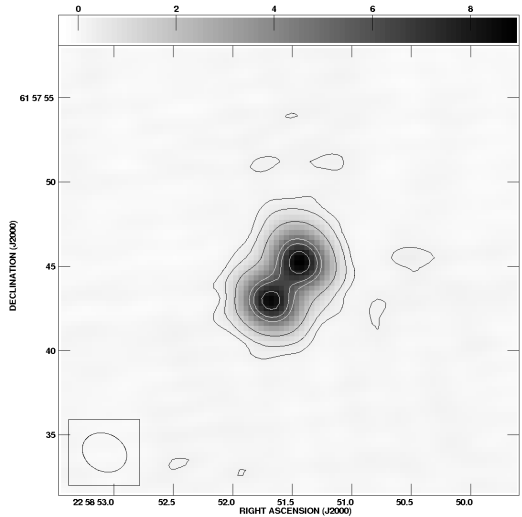}
\caption{Bipolar morphology observed at 8.4 GHz in IRAS 22568+6141. The contours are 60$\times$(-3, 3, 9, 27, 81, 100, 130)~$\mu$Jy/beam and the grey scale is in mJy/beam. The convolution beam is shown in the bottom left corner.}
\label{22568}
\end{figure}

Figure~\ref{22568} shows the double-peak morphology found in the target, which matches what observed at optical and near-IR wavelengths. The southern peak is not resolved in any of our maps, while the northern one is partly resolved with a size around $2.0''\times0.8''$ (PA$\sim50^\circ$). The emission measures are $12.3\times10^6$ and $14.8\times10^6$~cm$^{-6}$~pc, for the Southern and Northern lobe respectively, where the same size has been used for both blobs. 

If we assume a distance to the nebula of 6~kpc, from the emission measure we find  densities of $2.0\times10^4$~cm$^{-3}$ (North peak) and $1.8\times10^4$~cm$^{-3}$ (South peak). The corresponding decay times for electron recombination are $t_{decay}=\frac{7.6\times10^{4}}{(n_e/cm^{-3})}$~yr \citep{kwokbook}, then about 3.8 and 4.2 years for the N and S blobs respectively.
For the northern peak, which is partly resolved in our observations, we also calculate the ultraviolet luminosity necessary to produce the observed radio flux density as done for IRAS~18062+2410. In this case, we calculate the luminosity necessary if the star were at the centre of the blob, assuming the emitting volume is a sphere whose radius equals the geometric mean of the semiaxes  from the Gaussian fit. We obtain $Q\sim4.7\times10^{46}$~photons~s$^{-1}$. 
Because of the geometry of the source, this is to be regarded as a lower limit to the stellar UV luminosity, since the photons ionising the blob are only a fraction of all those emitted by the central star, and many may also be absorbed by circumstellar dust. By interpolation of the values in \citet{panagia} for class III, this gives us a lower limit of 24600 K for the temperature of the central star, slightly larger than what estimated by \citet{garcialario}. This lower limit points out a possible rapid increase of the temperature of the central star.

Our observations confirm that this object has started its PN evolutionary phase.

\section{Conclusions}
The transition from post-AGB stars to PN is a crucial point of the evolution of intermediate-mass stars. The development of an ionisation front determines major changes in the circumstellar environment and may contribute to the shaping of these intriguing sources.
Winds and jets are among the agents responsible for the early shaping of the nebulae during the early post-AGB phase, if not even before.

The ionisation of the envelope gives us the possibility to study the action of these agents around the star. Observations at optical wavelengths have pointed out that hot post-AGB stars show photometric variability, which in some instances has a quasi-periodic pattern \citep{arkhipova}.

Within our sample of pre- and young PN, we have found that 4 of our targets exhibit radio-continuum variability. In IRAS~18062+2410, the variability consists in a steadily increasing flux density, while in IRAS~17423-1755 it may follow a periodic pattern. The remaining two variable sources, namely IRAS~17516-2525 and 22568+6141, have both decreased their flux densities of about 10\% per year over a 4-year time. The cause of such a decrease is not clear. They may follow a variability pattern similar to IRAS~17423-1755, but in this source the variability occurs in a very compact region around the central star, while in IRAS~22568+6141 it involves two regions likely $\sim$1.5$''$  away from the star ($1.3\times10^{17}$~cm at 6~kpc). 

In near-IR images, the central star in IRAS~22568+6141 appears completely obscured, which implies  the presence of a thick circumstellar region of dust and molecular gas. Even at cm wavelengths, we do not detect any central emission, although this may be an angular resolution issue. We can speculate that this obscuring region is in fact a circumstellar disk and the two radio blobs are due to the action of a collimated wind onto the circumstellar nebula. In this scenario, if the wind slows down, fewer photons per second will reach the CSE. Because of the high density, the decay time is short and then the radio flux will decrease, as a result of recombinations. In an alternative  scenario, we can imagine that the central star is rapidly increasing its temperature, producing more and more photons at shorter wavelengths. In the blackbody approximation, the radiation from the central star peaks around 1207~\AA, if T$_{\mathrm{eff}}$=24000~K. Down to about 600~\AA~for silicates and 800~\AA~for amorphous carbon, the dust absorption coefficient increases with decreasing wavelength, therefore the hardening of the radiation from the central star would suffer stronger dust absorption determining a decrease of the ionization degree and of the observed radio flux density, until the radiation manages to destroy the dust grains. A decrease in radio continuum emission over a few years might then indicate a fast increase in stellar temperature. 

Variability may be present in several other targets in the sample, although the data do not allow for any conclusive statement.
Similar results have been obtained for other stars transiting between the late post-AGB and the PN phases, such as CRL~618 \citep{sanchez} and SAO~244567 \citep{sao}. 

While a linear increase in time of the flux density points out to the progression of the ionisation front in the envelope, quasi-periodic patterns may indicate the presence of jets or instabilities of the stellar wind (i.e., shocks).
Several models invoke instabilities and episodic ejections, or quasi-periodic variability. The MHD simulations in \citet{garcia} show that the collimated outflow can be subject to instabilities and form blobs (i.e., the knots observed in pre-PN), or an episodic jet can be produced by periodic variations of the magnetic field. Such instabilities typically have much longer time-scales than what observed in our targets and cannot be linked directly to our (relatively) short-term variations. 

\label{lastpage}

\section*{Acknowledgments}
The National Radio Astronomy Observatory is a facility of the National Science
Foundation operated under cooperative agreement by Associated
Universities, Inc.

\end{document}